\documentclass[aps,english,prl,floatfix,amsmath,superscriptaddress,tightenlines,twocolumn]{revtex4-2}
\usepackage{mathtools, amssymb}
\usepackage{graphicx}
\usepackage{intcalc}
\usepackage{mathrsfs}
\usepackage{physics}
\usepackage[caption=false]{subfig}
\usepackage[shortlabels]{enumitem}
\usepackage{soul}
\usepackage[utf8]{inputenc}
\usepackage{xcolor}
\usepackage{amsthm}
\usepackage{tikz}
\usetikzlibrary{decorations.pathreplacing,calligraphy}
\usepackage{bm}
\usetikzlibrary{patterns}
\usetikzlibrary{decorations.pathreplacing}
\usepackage{float}
\usepackage{comment}
\usepackage{enumitem}
\usepackage{tcolorbox}
\usepackage{hyperref}
\hypersetup{colorlinks=true,linkcolor=red,citecolor=magenta,urlcolor=blue}

\usepackage{algorithm}
\usepackage[noend]{algpseudocode}

\newcommand{\synd}{\text{synd}}
\newcommand{\data}{\text{data}}

\newcommand{\cS}{\mathcal{S}}

\newcommand{\IK}[1]{{\color{magenta}\footnotesize{(IK) #1}}}
\makeatother
\newcommand{\MF}[1]{{\color{blue}\footnotesize{(MF) #1}}}
\makeatother    
\newcommand{\AT}[1]{{\color{orange}\footnotesize{(AT) #1}}}

\makeatother

\usepackage[all]{hypcap}

\begin{document}

\title{Experimental demonstration of the advantage of adaptive quantum circuits}

\author{Michael Foss-Feig}
\affiliation{Quantinuum, 303 S. Technology Ct., Broomfield, Colorado 80021, USA}

\author{Arkin Tikku}
\affiliation{Centre for Engineered Quantum Systems, University of Sydney, Sydney, New South Wales 2006, Australia}

\author{Tsung-Cheng Lu}
\affiliation{Perimeter Institute for Theoretical Physics, Waterloo, Ontario N2L 2Y5, Canada}

\author{Karl Mayer}
\affiliation{Quantinuum, 303 S. Technology Ct., Broomfield, Colorado 80021, USA}

\author{Mohsin Iqbal}
\affiliation{Quantinuum, Leopoldstrasse 180, 80804 Munich, Germany}

\author{Thomas M. Gatterman}
\author{Justin A. Gerber}
\author{Kevin Gilmore}
\author{Dan Gresh}
\author{Aaron Hankin}
\author{Nathan Hewitt}
\author{Chandler V. Horst}
\author{Mitchell Matheny}
\author{Tanner Mengle}
\author{Brian Neyenhuis}
\affiliation{Quantinuum, 303 S. Technology Ct., Broomfield, Colorado 80021, USA}

\author{Henrik Dreyer}
\affiliation{Quantinuum, Leopoldstrasse 180, 80804 Munich, Germany}

\author{David Hayes}
\affiliation{Quantinuum, 303 S. Technology Ct., Broomfield, Colorado 80021, USA}

\author{Timothy H. Hsieh}
\affiliation{Perimeter Institute for Theoretical Physics, Waterloo, Ontario N2L 2Y5, Canada}

\author{Isaac H. Kim}
\affiliation{Department of Computer Science, UC Davis, Davis, CA 95616, USA}

\begin{abstract}
Adaptive quantum circuits employ unitary gates assisted by mid-circuit measurement, classical computation on the measurement outcome, and the conditional application of future unitary gates based on the result of the classical computation. In this paper, we experimentally demonstrate that even a noisy adaptive quantum circuit of constant depth can achieve a task that is impossible for \emph{any} purely unitary quantum circuit of identical depth: the preparation of long-range entangled topological states with high fidelity. We prepare a particular toric code ground state with fidelity of at least $76.9\pm 1.3\%$ using a constant depth ($d=4$) adaptive circuit, and rigorously show that no unitary circuit of the same depth and connectivity could prepare this state with fidelity greater than $50\%$.

\end{abstract}
\maketitle

\emph{Introduction ---} Quantum computers are capable of efficiently solving problems that are likely intractable for classical computers~\cite{Feynman1982,Shor1997}, and rapid advances in quantum computing technology 
has prompted significant interest in using near-term quantum computers to solve problems of practical interest~\cite{Preskill2018}. A key challenge in this pursuit is that operations in today's quantum computers are inevitably accompanied by occasional errors; if allowed to accumulate uncorrected, these errors will corrupt the output of a computation. It is widely expected that quantum error correction will ultimately tame such errors, but in the meantime, attempts to achieve quantum advantage on problems of practical interest have hinged on efforts to partially mitigate the impact of errors, rather than eliminate them. Since errors accompany quantum operations, the most obvious strategy to reduce their impact is to reduce the number of quantum operations required for a given algorithm.

\begin{figure}[t!]
\begin{center}
\includegraphics[width=1.0\columnwidth]{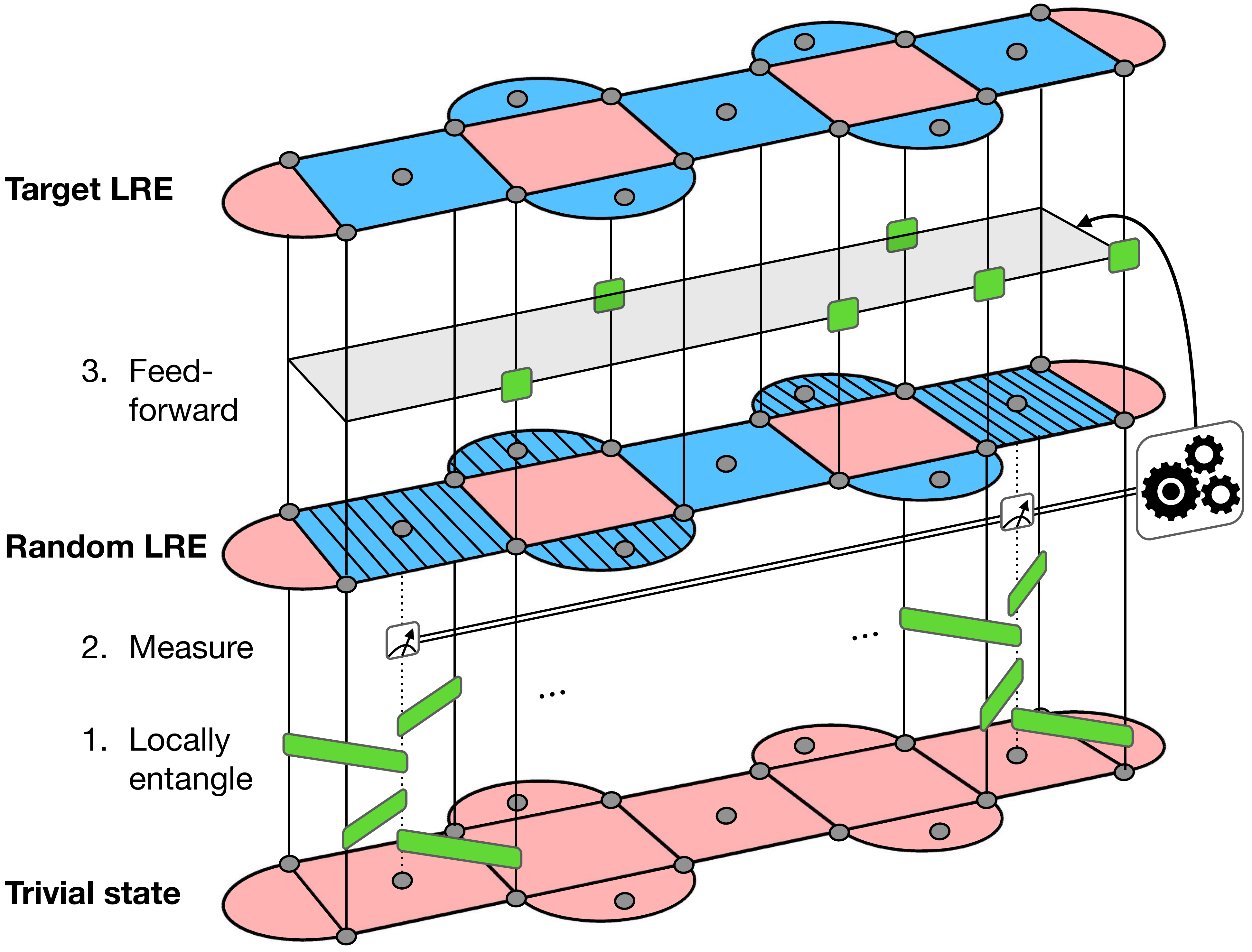}
    \caption{Local and finite-depth entangling operations---when augmented with local measurements--- can generate arbitrarily large but random long-range entangled (LRE) states conditioned on the random measurement outcomes.  Feed-forward, in this case single-qubit gates conditioned on the measurement outcomes, can then steer this random state to a desired target state with unconditional long-range entanglement.}
\label{fig:adaptive_circuits}
\end{center}
\end{figure}

One appealing approach to reducing quantum resources is to use \emph{adaptive quantum circuits}. These are quantum circuits that are assisted by mid-circuit measurement, classical computation on those measurement outcomes, and feed-forward to future quantum operations based on the result of the computation (Fig.\,\ref{fig:adaptive_circuits}). While the measurement, classical computation, and feed-forward can all in principle be transformed into a sequence of unitary gates, there are numerous potential reasons to prefer an adaptive quantum circuit. Unlike quantum computation, classical computation is essentially noise-free, more readily available, not geometrically constrained by qubit layout, and more easily scalable.  Especially in quantum technologies with long natural time scales---such as qubits based on trapped ions or neutral atoms---classical computation can also be extremely fast compared to quantum operations, and can therefore be used to reduce the total run-time and associated memory errors of quantum computations. Thus, by leveraging fast and accurate classical computational resources, one can hope to retain the advantage of using a quantum computer while minimizing the requisite quantum resources.

It has been noted that adaptive quantum circuits can significantly outperform circuits without classical feed-forward in some tasks, such as preparation of ground states of many-body quantum systems. For example, it is well-known that certain highly entangled many-body ground states, such as the topologically ordered ground states of Kitaev's toric code~\cite{Kitaev2003}, cannot be prepared using a quantum circuit whose depth is a constant that is independent of the system size~\cite{Bravyi2006}. However, by measuring a subset of qubits and conditioning future unitary operations on the measurement outcomes, it is possible to prepare the ground state exactly in constant depth~\cite{Raussendorf2003,Raussendorf2005}, thereby reducing the idle time over which quantum coherence must be maintained. This observation was recently extended to a larger class of quantum many-body ground states~\cite{Verresen2021,Tantivasadakarn2021,Bravyi2022,Lu2022,Tantivasadakarn2022,Tantivasadakarn2022a}. These theoretical results firmly establish that constant-depth adaptive quantum circuits can prepare quantum states that are impossible to prepare using constant-depth quantum circuits without feed-forward.  However, experimental demonstrations of this fact have proved difficult to achieve. The primary technical challenge in implementing adaptive circuits is the requirement to perform partial measurements of a subset of qubits in the middle of a quantum circuit with minimal cross-talk on unmeasured qubits, return those results to a classical computer for processing, and then condition future operations on the results of that processing in real time.  On a physical level, high-fidelity measurement is an intrinsically slow and destructive process, at odds with the requirement that parts of the quantum computer remain well-isolated and coherent during the measurement process. And, depending on the architecture, processing of classical information much faster than the coherence time can be challenging \cite{skoric2022parallel}. 

While there has been remarkable progress towards implementing adaptive circuits in numerous quantum computing platforms (e.g., Refs.\,\cite{Barrett:2004vw,Pfaff:2013ue,Riste:2013um,Wan875}), only recently have hardware developments in superconducting qubits and trapped ions made it possible to run adaptive quantum circuits on large-scale, universal quantum computers \cite{pino_2021,PhysRevX.11.041058,PhysRevLett.127.100501}. In this manuscript, we show that hardware has now reached a stage where adaptive quantum circuits are not only possible, but can outperform unitary circuits: Using the Quantinuum H1-1 quantum computer, we experimentally demonstrate a quantifiable and verifiable advantage in using an adaptive quantum circuit over any purely unitary quantum circuit of identical depth when both circuits employ gate sets restricted to the same geometry. We propose a concrete metric, which can be estimated efficiently on a quantum computer, that can unambiguously establish the difference in resource requirements between circuits with and without feed-forward.

\begin{figure}[!t]
\centering
\includegraphics[width=1.0\columnwidth]{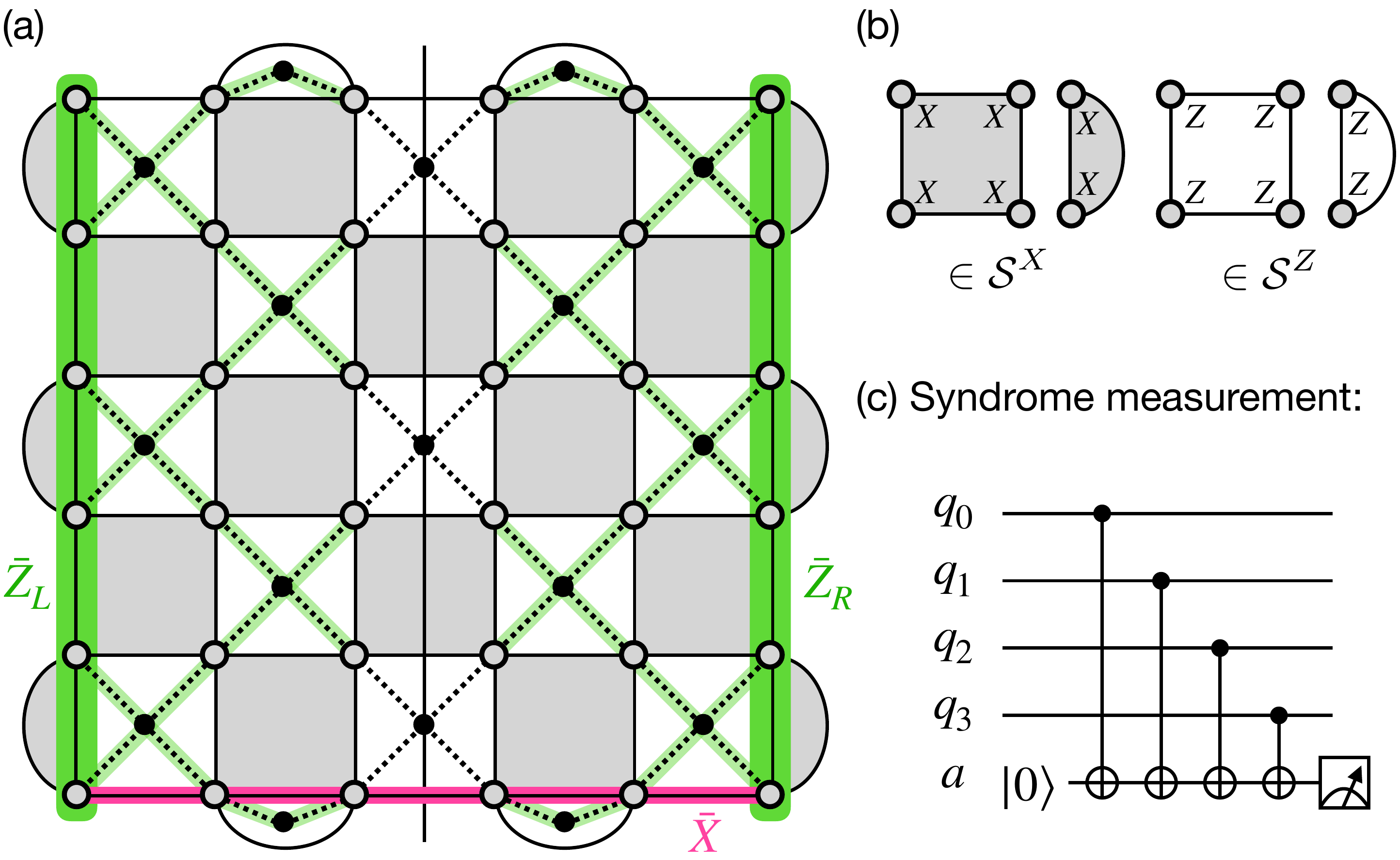}
\label{fig:setup}
\caption{
(a) 6x6 rotated surface code. $X$ and $Z$ type stabilizers live on shaded and unshaded plaquettes, respectively, and are defined as produces of Pauli matrices on the plaquette vertices as in (b). Ancilla-assisted measurements of the $Z$ stabilizers, shown in (c), define a geometry amongst the combined data and ancilla qubits shown as dashed lines in (a).}
\end{figure}

More specifically, we consider the task of preparing a certain ground state of Kitaev's toric code~\cite{Kitaev2003,Bravyi1998} with open boundary conditions, using a quantum circuit of depth $d=4$ \footnote{We use the fairly standard definition that depth is the minimum number of required layers of two-qubit gates such that no layer contains gates having shared-support.  Other definitions, for example in which single-qubit gates or measurement are included, would slightly change the code distance requirements for a 50\% upper bound in the absence of feed-forward, but would not appreciably change the arguments made in this paper.} and specific geometric constraints (defined by the requirements of making ancilla-based parity measurements, see Fig.\,\ref{fig:setup}). We prove an upper bound on the achievable fidelity between the ground state and any state prepared by such a constant depth quantum circuit on the same geometry and without feed-forward, which is $50\%$. We then experimentally prepare the ground state with fidelity of $76.9 \pm 1.3 \%$ using an adaptive circuit with the same depth. Our experiment thus firmly establishes that, given access to the same amount of quantum computational resources with respect to available gates and circuit depth, adaptive quantum circuits can perform tasks that are impossible for quantum circuits without feed-forward.

\emph{Task ---} We compare two methods for preparing the logical $\ket{\bar{+}}$ state of the rotated surface code (RSC) (throughout the paper, states/operators with bars over them should be understood as logical states/operators). The first method utilizes an adaptive circuit based on stabilizer measurements and corrections conditioned on the measurement outcomes \cite{PhysRevA.86.032324}.  The measurements project an initially uncorrelated state into a random eigenstate of the Hamiltonian
\begin{align}
\label{eq:Ham}
H=-\sum_{S\in\mathcal{S}^X}S-\sum_{S'\in\mathcal{S}^Z}S',
\end{align}
Here, $S\in \mathcal{S}^X$ ($S'\in \mathcal{S}^Z$) is a stabilizer formed by multiplying together all $X$($Z$) Pauli operators on the vertices of a given plaquette shown in Fig.\,\ref{fig:setup}(a,b) (shaded plaquettes are associated with $X$ stabilizers, and unshaded plaquettes with $Z$ stabilizers). Corrections based on the measurement outcomes then steer this eigenstate into the two-fold degenerate ground state manifold (code space), specifically to the logical $\ket{\bar{+}}$ state [the unique state in the ground-state manifold with $+1$ eigenvalue in the logical $\bar{X}$ operator shown in Fig.\,\ref{fig:setup}(a)]. All stabilizers in $\mathcal{S}^X$ can be satisfied by choosing an initial product state of all qubits in the $\ket{+}$ state. Therefore we will only need to measure stabilizers in $\mathcal{S}^Z$, which we do by placing an ancilla qubit inside each associated plaquette and applying controlled-NOT gates targeting the ancilla and controlled on the data qubits at the vertices of the plaquette [Fig.\,\ref{fig:setup}(c)].  We take this pattern of gates, shown as dashed lines in Fig.\,\ref{fig:setup}(a), to \emph{define} the geometry of the model, i.e.\ gates are allowed to act between any qubits connected directly by gates during the adaptive state preparation protocol. Note that this preparation method requires a circuit of constant depth $d=4$, in the sense that one could grow the code arbitrarily large in either the vertical or horizontal direction and still prepare the logical $\ket{\bar{+}}$ state in depth $d=4$. The second method is non-adaptive, and restricted to use two-qubit gates that are local in the geometry as previously defined (here and elsewhere, we use the term \emph{local unitary} to refer to spatial locality in this geometry). We do not consider a specific encoding circuit, but rather would like to make statements about \emph{all} such local unitary circuits.

\begin{figure}[!t]
\centering
\includegraphics[width=0.9\columnwidth]{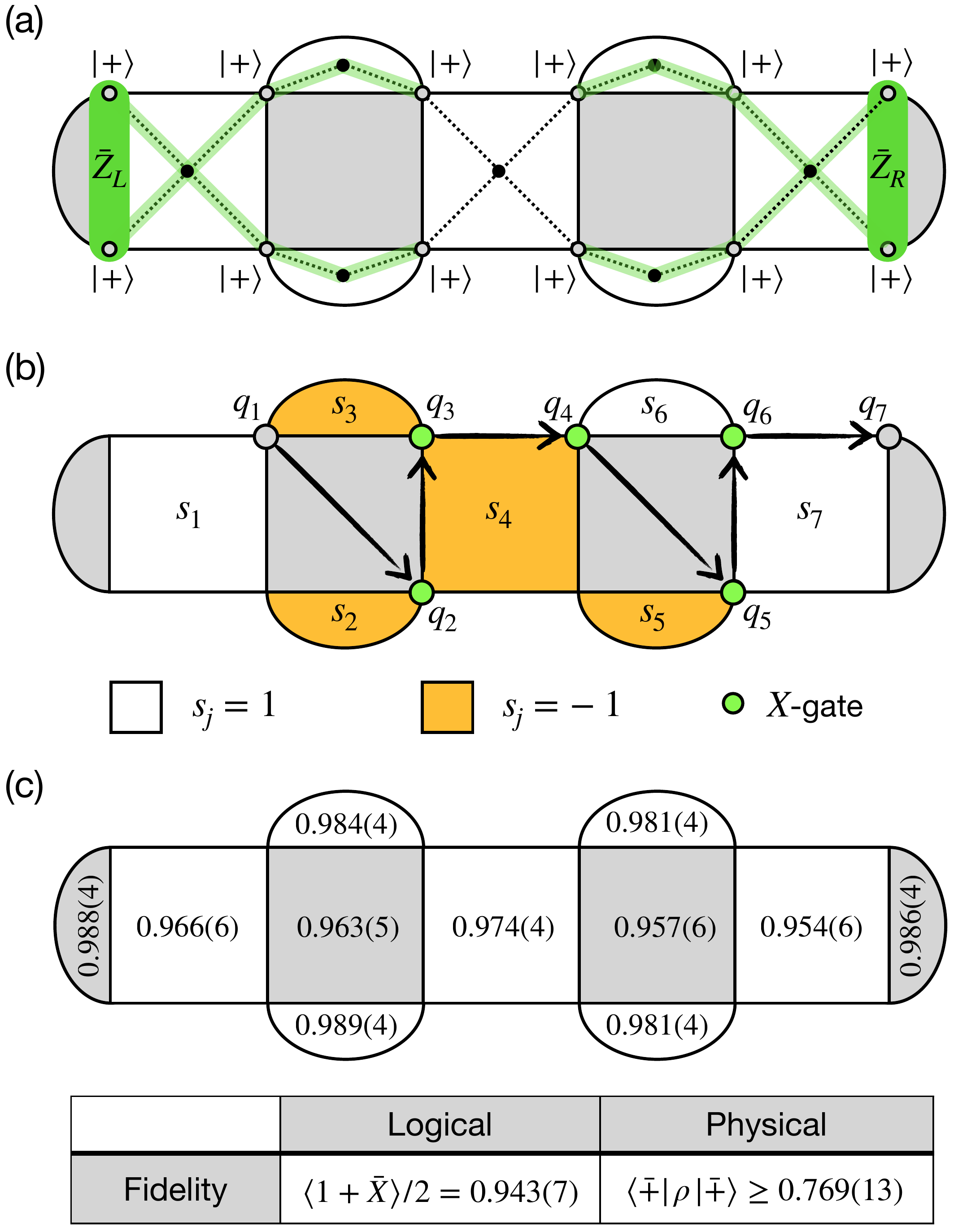}
\label{fig:state_prep}
\caption{Finite-depth preparation of a long-range entangled state using local unitaries plus measurement and classical feed-forward. (a) An elongated surface code is initialized by putting all data qubits in the $\ket{+}$ state and then measuring Z-type stabilizers using a depth-4 circuit.  Note that the light cones of $Z_l$ and $Z_r$ induced by a local depth-4 unitary (green) are non-overlapping. (b) Z-plaquette defects (orange) are purged by applying $X$-corrections along the data-qubit chain shown. (c) Stabilizer-resolved and global fidelities. The value in each plaquette reports the expectation value $\langle 1+S\rangle/2$ for the associated stabilizer $S$.}
\end{figure}

In what follows we first show that for a sufficiently large code, no unitary circuit under these connectivity constraints and with depth $d\leq 4$ can approximate the logical $\ket{\bar{+}}$ state with fidelity exceeding $50 \%$. We then describe the implementation of the adaptive state preparation procedure in detail for the elongated $2\times 6$ strip of the RSC shown in Fig.\,\ref{fig:state_prep}, as well as an efficient experimental method of lower bounding the fidelity of the prepared state.

\emph{Upper bound on fidelity from local unitaries---} The surface code logical states are long-range entangled, and one consequence is that they have large correlations between physical degrees of freedom spanning the code.  To see this, note that the logical $\bar{Z}$ operator has many equivalent physical representations, two of which are located at either the left or right boundary of the code, denoted $\mathscr{B}_L$ or $\mathscr{B}_R$, respectively [Fig.\,\ref{fig:setup}(a)]. We call these two physical representations $\bar{Z}_{L/R}=\prod_{j\in \mathscr{B}_{L/R}}Z_j$. Within the code space both representations act as logical $Z$, and hence their product acts as the logical identity.  As a result the logical $\ket{\bar{+}}$ state, which has vanishing expectation value in any representation of $\bar{Z}$, has maximal connected correlation functions between these two physically distant representations,
\begin{align}
\langle \bar{Z}_L\bar{Z}_R\rangle_c\equiv \bra{\bar{+}}\bar{Z}_L\bar{Z}_R\ket{\bar{+}}-\bra{\bar{+}}\bar{Z}_L\ket{\bar{+}}\bra{\bar{+}}\bar{Z}_R\ket{\bar{+}}=1.
\end{align}

Assuming an initial product-state $\ket{\psi_i}$, it is straightforward to show that two operators $A$ and $B$ will have vanishing connected correlations in the final state $\ket{\psi_f}=U\ket{\psi_i}$ unless their past causal cones induced by the unitary $U$ have overlapping support. Suppose we have some constant depth local unitary $U$ that prepares $\ket{\bar{+}}$ from an initially uncorrelated state.  The past causal cones of $\bar{Z}_L$ and $\bar{Z}_R$ have finite size due to the finite depth and locality of $U$, and therefore as we scale up the code distance we will eventually find that the connected correlation function $\langle \bar{Z}_L\bar{Z}_R\rangle_c$ vanishes, contradicting the assumption that we have prepared the logical $\ket{\bar{+}}$ state. One can similarly show that no state within the code space can be created by a constant-depth local unitary; the constant-depth approach using measurement and feed-forward therefore achieves something not possible by purely unitary operations.

In an actual experiment, when we attempt to prepare the $\ket{\bar{+}}$ state by \emph{any} method we will inevitably have errors, achieving only a noisy approximation $\rho\approx \ket{\bar{+}}\bra{\bar{+}}$.  It is natural to ask: Given a noisy approximation, how good must the approximation be in order to rule out having prepared it by a local unitary in constant depth? 
%
%
To answer this question, we can upper bound the fidelity of the state $\rho$ with respect to the ideal state $\ket{\bar{+}}$ in terms of their respective expectation values in an arbitrary POVM $\{M_j\}$.  Denoting the expectation values of $M_j$ in the ideal and noisy states as $p_j={\rm Tr}(\ket{\bar{+}}\bra{\bar{+}}M_j)$ and $q_j={\rm Tr}(\rho M_j)$, respectively, then from the basic properties of fidelities of quantum states \cite{wilde2011classical} we have
\begin{align}
F\equiv\bra{\bar{+}}\rho\ket{\bar{+}}\leq \Big(\sum_{j}\sqrt{p_j q_j}\Big)^2.
\end{align}
We choose the POVM elements to be products of projectors onto the positive and negative eigenvalue subspaces of $\bar{Z}_{L/R}$,
\begin{align}
M_j\in\{\pi_L^{+}\pi_R^{+},~\pi_L^{+}\pi_R^{-},~\pi_L^{-}\pi_R^{+},~\pi_L^{-}\pi_R^{-}\},
\end{align}
where $\pi_{L/R}^{\pm}=(1\pm\prod_{j\in \mathscr{B}_{L/R}}Z_j)/2$. By assumption of the finite-depth preparation and resulting lack of correlations between $\bar{Z}_L$ and $\bar{Z}_{R}$, the POVM expectation values for the generated state must have the form \footnote{This statement holds true even for a noisy version of the circuit, as long as the noise channel associated with a given gate is supported on the same qubits the gate is.}
\begin{align}
\vec{q}=\big(ab,a(1-b),(1-a)b,(1-a)(1-b)\big).
\end{align}
Meanwhile the target state is maximally correlated, and $\vec{p}=(1,0,0,1)/2$.  Defining $\vec{u}=
(\sqrt{a},\sqrt{1-a})$ and $\vec{v}=(\sqrt{b},\sqrt{1-b})$, we find the following upper bound on the fidelity of any state achieved by a depth-4 local unitary
\begin{align}
F\leq\frac{1}{2}(\vec{u}\cdot\vec{v})^2\leq 1/2.
\end{align}
Note that this bound applies for any fixed depth once the code distance is sufficiently large that the maximal past-causal cones of $Z_L$  and $Z_R$ are non-overlapping. The depth of the adaptive preparation protocol, $d=4$, leads to light cones depicted via the green shading in Figs.\,\ref{fig:setup}(a) and Fig. \ref{fig:state_prep}(a), which do not overlap for a surface code with horizontal length of 6 or greater. 

\emph{Adaptive state preparation protocol ---}\label{sec:adaptive_state_prep_protocol}
We will now describe an experimental violation of this upper bound by using an adaptive circuit. Let $\cS^{Z}(\cS^{X})$ denote the set of $Z$-type ($X$-type) stabilizers and let $\bar{Z}(\bar{X})$ denote the logical $Z$($X$) operator. We aim to prepare the logical state $\ket{\bar{+}}$, which is the unique state stabilized by any element of the group generated by $\cS^{Z}\cup\cS^{X}\cup \{\bar{X}\}$.  We first prepare an initial state consisting of all data qubits individually in the $\ket{+}$ state [Fig.\,\ref{fig:state_prep}(a)], which is stabilized by any operator in $\cS^{X}\cup \{\bar{X}\}$ but not by $\bar{Z}$. We then measure each $Z$-type stabilizer using the ancilla pertaining to the associated plaquette, randomly obtaining values of $\pm 1$ for each stabilizer [Fig.\,\ref{fig:state_prep}(b)]; since all measured $Z$-type stabilizers commute with all operators in $\cS^{X}\cup\{\bar{X}\}$, the post-measurement state is still stabilized by them.  If all $Z$-stabilizer measurements were $+1$ [white stablizers in Fig.\,\ref{fig:state_prep}(b)] then we are done, as the state is also now stabilized by any element in $\cS^{Z}$, however the probability of obtaining this outcome is exponentially small in the number of $Z$-type stabilizers.  Fortunately the $-1$ defects for $Z$-type-stabilizer outcomes [orange stabilizers in Fig.\,\ref{fig:state_prep}(b)] can be freely modified without changing the state's eigenvalues associated with any operators in $\cS^{X}\cup\{\bar{X}\}$ by applying single qubit Pauli-$X$ gates conditioned on the stabilizer measurement outcomes.


We carry out the state preparation protocol on the Quantinuum H1-1 quantum computer using 19 qubits (12 data qubits and 7 ancilla qubits). In addition to the flexibility of implementing any geometry without the overhead of logical SWAPS (thanks to arbitrary dynamic repositioning of ions during a circuit), Quantinuum's H-Series hardware supports a variety of real-time classical logic \cite{ryan2022implementing}, including low-level logical primitives supported in an extended version of OpenQASM 2.0 \cite{PhysRevX.11.041058}.  We write the stabilizer measurements into a classical register $s$ (with bit labels $s_j=1,...,7$), with the ordering shown in Fig.\,\ref{fig:state_prep}(b)].  We then loop through the qubits $q_j$ along that path, applying an $X$ gate (and updating the stabilizer values $s_j$ of impacted stabilizers) whenever the left-most of the two stabilizers containing that qubit has the wrong value, ultimately purging all defects by the end of the path. 

\emph{Fidelity estimation---}The $\ket {\bar{+}}$ state is the unique state with $+1$ eigenvalue for all stabilizer generators in $\mathcal{S}^X\cup\mathcal{S}^Z\cup\{\bar{X}\}$. Therefore the fidelity of any state $\rho$ with respect to $\ket{\bar{+}}$ is given by the expectation value of a product of projectors into the $+1$ eigenspace of all generators,
\begin{align}
F&=\Big\langle \Big( \frac{1+\bar{X}}{2}\prod_{S\in S^X}\frac{1+S}{2} \Big)\Big(\prod_{S'\in\mathcal{S}^Z}\frac{1+S'}{2} \Big)\Big\rangle\\
&\equiv\langle P_x P_z\rangle
\end{align}
where $\langle\star\rangle={\rm Tr}(\rho \star)$ and $P_{x(z)}$ denote the first(second) parenthetical term in the line above.  Since $P_{x(z)}$ exclusively contains Pauli $X$($Z$) operators, measuring all qubits in either the uniform $x$ or uniform $z$ basis provides an estimate of either $\langle P_x\rangle$ or $\langle P_z\rangle$.  Then we can bound the fidelity as
\begin{align}
F &=\Big\langle\Big[1-(1-P_x)\Big]\Big[1-(1-P_z)\Big]\Big\rangle\nonumber\\
&\geq \langle P_x\rangle+\langle P_z\rangle-1,\label{eq:fidelity_bound}
\end{align}
where the second line follows from dropping the correlation function between $x$ and $z$ outcomes, $\langle(1-P_x)(1-P_z)\rangle$, which is non-negative because $1-P_x$ and $1-P_z$ commute and are positive semi-definite. We estimate the fidelity by performing the adaptive state preparation procedure $1000$ times followed by $X$ basis measurement, and another $1000$ times followed by $Z$ basis measurements. The data provides direct estimates of $\langle P_x\rangle$ and $\langle P_z\rangle$, from which we infer a fidelity lower bound $F\geq 0.769(13)$ using Eq.\,(\ref{eq:fidelity_bound}).  Fidelities of the individual stabilizers, along with the logical fidelity $\langle 1+\bar{X}\rangle/2$, are reported in Fig.\,\ref{fig:state_prep}(c).  All error bars are $1\sigma$ uncertainties obtained from bootstrap resampling of the data (100 resamples).

\emph{Summary and Outlook ---} We have experimentally demonstrated preparation of a toric code ground state using a depth-4 adaptive circuit involving measurement and feed-forward, and we have shown that the fidelity achieved is not possible using a local unitary circuit of the same depth, illustrating the significant potential of adaptive circuits for efficiently preparing long-range entangled states. Similar conclusions regarding the inadequacy of constant-depth local unitaries apply to more local metrics as well, such as the energy density~\cite{Anshu2022,Tikku2022,Anshu2022a}. It is expected that even upon relaxing geometric locality, $k$-local unitary circuits (those comprised of gates that each act non-trivially on at most $k$ qubits) also cannot achieve high fidelity with respect to a topologically ordered state in constant depth. The required depths for exact $k$-local unitary preparation of topologically ordered states are logarithmic in code distance \cite{Aharonov2018}, as opposed to linear for geometrically local unitaries \cite{Bravyi2006}, and thus substantially more qubits are likely needed to exceed the fidelity achievable by any $k$-local depth-4 circuit \footnote{There is already a logarithmic lower bound for exact state preparation~\cite{Aharonov2018}, but a generalization of this bound to a general non-unity fidelity is not known to us.}.

It would be interesting to implement other adaptive protocols for preparing a wider class of long-range entangled states, including quantum critical points and non-abelian topological order.  Furthermore, even for preparing a given topologically ordered state such as a ground state of the toric code, there are several alternative adaptive protocols.  While they are all constant depth, they may have relative advantages depending on the hardware details, and it would be useful to quantify these differences through simulation and implementation.    

\begin{acknowledgments}
This work was made possible by a large group of people, and the authors would like to thank the entire Quantinuum team for their many contributions. MF, KM, and DH thank Matt DeCross, Ciaran Ryan-Anderson, Natalie Brown, Reza Haghshenas, and Eli Chertkov for helpful discussions, and Russell Stutz and Chris Langer for helpful comments on the manuscript. IK thanks Anurag Anshu for helpful discussions. AT would like to thank Thomas B. Smith for helpful comments. AT was supported by the Sydney Quantum Academy, Sydney, NSW, Australia. T-C.L. and T.H.'s research at Perimeter Institute is supported in part by the Government of Canada through the Department of Innovation, Science and Economic Development Canada and by the Province of Ontario through the Ministry of Colleges and Universities.  The experimental data in this work was produced by the Quantinuum H1-1 trapped ion quantum computer, Powered by Honeywell.
\end{acknowledgments}

\end{document}